
%
\input phyzzx
%
%
%
%
\newcount\lemnumber   \lemnumber=0
\newcount\thnumber   \thnumber=0
\newcount\conumber   \conumber=0

\def\myeq{{\rm \chapterlabel.\the\equanumber}}

\def\Lemma{\par\noindent\global\advance\lemnumber by 1
           {\bf Lemma\ (\chapterlabel\the\lemnumber)}}
\def\Corollary{\par\noindent\global\advance\conumber by 1
           {\bf Corollary\ (\chapterlabel\the\conumber)}}
\def\Theorem{\par\noindent\global\advance\thnumber by 1
           {\bf Theorem\ (\chapterlabel\the\thnumber)}}

%
%
\def\e{\adveq\eqno{\rm (\chapterlabel\the\equanumber)}}
\def\adveq{\global\advance\equanumber by 1}


%
%
\font\tensl=cmsl10
\font\tenss=cmssq8 scaled\magstep1
\outer\def\quote{
   \begingroup\bigskip\vfill
   \def\endquote{\endgroup\eject}
    \def\par{\ifhmode\/\endgraf\fi}\obeylines
    \tenrm \let\tt=\twelvett
    \baselineskip=10pt \interlinepenalty=1000
    \leftskip=0pt plus 60pc minus \parindent \parfillskip=0pt
     \let\rm=\tenss \let\sl=\tensl \everypar{\sl}}
\def\from#1(#2){\smallskip\noindent\rm--- #1\unskip\enspace(#2)\bigskip}

\def\CALT{\address{Division of Physics, Mathematics
and Astronomy\break
Mail Code 452--48\break
California Institute of Technology\break
Pasadena, CA 91125}}

\def\r#1{$\lb \rm#1 \rb$}

%
%
\def\rarrow{\rightarrow}

\def\semidirect{\mathrel{\raise0.04cm\hbox{${\scriptscriptstyle |\!}$
\hskip-0.175cm}\times}}


\def\ref#1{$^{#1}$}

\def\half{{1\over2}}
\def\lb{\lbrack}
\def\rb{\rbrack}

\def\diam{{\hbox{\hskip-0.02in
\raise-0.126in\hbox{$\displaystyle\bigvee$}\hskip-0.241in
\raise0.099in\hbox{ $\displaystyle{\bigwedge}$}}}}

\def\bw#1#2#3#4#5{{w\left(\matrix{#1&#2\cr#3&#4\cr}\bigg\vert #5\right)}}

\def\equiv{\sim}
\date{May, 1993}
\titlepage
\title{On Solvable Lattice Models and Knot Invariants}
\author{Doron Gepner}
\CALT\foot{On leave from the Weizmann Institute, Israel. Incumbent
of the Sorrela and Henry Shapiro Chair.}
\vskip15pt
\abstract
Recently, a class of solvable interaction round the face lattice models
(IRF) were constructed for an arbitrary rational conformal field theory (RCFT)
and an arbitrary field in it. The Boltzmann weights of the lattice models
are related in the extreme ultra violet limit to the braiding matrices
of the rational conformal field theory. In this note we use these new
lattice models to construct a link invariant for any such pair of an RCFT
and a field
in it. Using the properties of RCFT and the IRF lattice models, we prove
that the invariants so constructed always obey the Markov properties, and
thus are true link invariants. Further, all the known link invariants,
such as the Jones, HOMFLY and Kauffman polynomials arise in this way,
along with giving a host of new invariants, and thus also a unified approach
to link polynomials. It is speculated that all link invariants arise
from some RCFT, and thus the problem of classifying link and knot
invariants is equivalent to that of classifying two dimensional conformal field
theory.
\endpage
The intriguing interplay between knot theory and two dimensional
physics has benefited considerably both fields (for a review, see, e.g.,
\REF\Wadati{M. Wadati, T. Deguchi and Y. Akutsu, Phys. Rep. 180 (1989) 247}
\r\Wadati). The purpose of this note is to put forwards a general framework
for link invariants stemming from solvable lattice models. It was recently
shown that solvable fusion interaction round the face (IRF) lattice models
are in a one--to--one correspondence with a pair of a rational conformal field
theory
and a field in it \REF\One{D. Gepner, Foundations of rational conformal field
theory, I, Caltech preprint, CALT 68--1825, November (1992)}\r\One.
It follows as we shall see that for each such pair
one can form a link invariant, and that this class of link invariants is
in a one to one correspondence with such pairs.

Let us review the construction of the Boltzmann weights described in ref.
\r\One.
Consider a rational conformal field theory (RCFT) $\cal O$, and a field in
it, $x$, which for simplicity we shall assume to be a primary field.
We than construct a solvable IRF model, denoted by IRF$({\cal O},x)$ following
\r\One, whose admissibility conditions are given by fusion with respect for
$x$ and whose Boltzmann weights reduce in the extreme ultra violet limit
to a specialization of the braiding matrix of the RCFT (see \r\One\ for more
detail). We put on the vertices of the lattice, which is a square two
dimensional one, state variables which
are the primary fields of $\cal O$ and are labeled by $a,b,c,\ldots$. The
pair $a$ and $b$ is allowed to be on the same link, $a\equiv b$, if and
only if, the fusion coefficient $N_{ax}^b>0$. The partition function
of the model is
$$Z=\sum_{\rm configurations} \prod_{\rm faces} \bw a b c d u,\e$$
where $a,b,c$ and $d$ are the four states (primary fields) on the vertices of
the
face, $\bw a b c d u$ is the Boltzmann weight associated to the
face, and $u$ is a spectral parameter which labels a family of models.
The Boltzmann weights obey the star triangle equation (STE), from which
it follows that the transfer matrices for different values of the
spectral parameter $u$ commute, and thus the model is solvable.

The Boltzmann weights of the model IRF$({\cal O}, x)$ were given in ref.
\r\One, and are conveniently described in an operator form.
To do so define the operator $X_s(u)$, the face transfer matrix, by
$$X_s(u)_{l_1,l_2,\ldots,l_n}^{m_1,m_2,\ldots,m_n}=
\prod_{i\neq s}\delta(l_i,m_i) \bw {l_{i-1}} {m_i} {l_i} {l_{i+1}} u,\e$$
where $l_i$ and $m_i$ are the states on two adjacent diagonals of the
lattice. The face transfer matrix of the model IRF$({\cal O},x)$ is \r\One
$$X_s(u)=\sum_{a=1}^N P^a_s f^a(u),\e$$
where $a=1,2,\ldots,N$ labels the fields appearing in the operator product
$x\cdot x$,
$P^a$ is a projection operator of the braiding matrix on the $a$ field
in the operator product defined by
$$P^a=\prod_{j=1\atop j\neq a}^N {B_s-\lambda_j\over \lambda_a-\lambda_j},\e$$
and where $B_s$ is the braiding matrix of the RCFT at the face $s$, and
$\lambda_j$ are its eigenvalues, which are given by
$$\lambda_j=e^{i\pi (2\Delta_x-\Delta_j)},\e$$
and $\Delta_x$ and $\Delta_j$ are the conformal dimensions of the field
$x$ and the $j$ field in the operator product $x\cdot x$, respectively.

The functions $f^a(u)$ are defined by,
$$f^a(u)=\prod_{j=1}^{a-1} \sin (\zeta_j+u) \prod_{j=a}^{N-1}
\sin(\zeta_i-u),\e$$
where
$$\zeta_i=\pi(\Delta_{i+1}-\Delta_i)/2,\e$$
and $\lambda=\zeta_1$ is the
crossing parameter of the model.
The projection operators obey,
$$\eqalign{P^a_s P^b_s&=\delta_{ab} P_s^a,\cr
           1&=\sum_{a=1}^N P_a,\cr
          B_s&=\sum_{a=1}^N P^a_s \lambda_a,\cr}\e$$
from which it follows that the face transfer matrix obeys the unitarity
condition,
$$X_s(u) X_s (-u)=\rho(u)\rho(-u),\e$$
where the unitarity factor is
$$\rho(u)=f^N(u)=\prod_{i=1}^{N-1} \sin(\zeta_i+u).\e$$
Also, this implies the regularity condition,
$$X_s(0)=\rho(0)\cdot 1.\e$$
An important, and highly non trivial, property of the Boltzmann weights
is the crossing symmetry,
$$\bw a b c d {\lambda-u}=\left({\psi_b \psi_c \over \psi_a \psi_d}
\right)^\half \bw c a d b u,\e$$
where the crossing multiplier $\psi_a$ is given in terms of the torus
modular function $S_{ab}$,
$$\psi_a={S_{a,0}\over S_{0,0}},\e$$
where `$0$' denotes the unit field.
Repeating the crossing transformation twice implies the charge conjugation
symmetry:
$$\bw a b c d u=\bw d c b a u.\e$$

It is convenient to define the two braiding operators,
$$G_i^\pm=\lim_{u\rarrow \pm \infty} X_i(u)/\rho(u),\e$$
where $G_i^+$ (denoted also for simplicity by $G_i$)
differs from the conformal braiding matrix $B_i$ by an
irrelevant phase. In terms of the Boltzmann weights, this is
\def\bs#1#2#3#4#5{\sigma\left(\matrix{ #1 & #2 \cr #4& #3\cr}\bigg\vert
#5\right)}
$$\bs a b c d {\pm}=\lim_{u\rarrow\pm\infty} \bw a b c d u/\rho(u),\e$$
from which it follows that $G_i^+=(G_i^-)^\dagger$, i.e., they are complex
conjugates of each other, and that $G_i^\pm$ obey the Braid group relationships
which are
$$\eqalign{G_i G_{i+1} G_i&=G_{i+1} G_i G_{i+1},\cr
           G_i G_j&=G_j G_i \qquad {\rm for\ } |i-j|>1,\cr }\e$$
which is the relation obeyed by the generators of the braiding group,
i.e., $G_i$
can be considered as the generator of the braiding of the $i$ and
$i+1$ strands in a braid. By Artin theorem these are the generating
relations for the braid group.

A link is formed by connecting the end points of a braid. Labeling
the end points $l_1,l_2,\ldots ,l_n$ and $m_1,m_2,\ldots,m_n$, as
before, we connect with a strand the $l_i$ and $m_i$ end points, for
all $i$. This procedure is ambiguous as different braids may give
the same (topologically) link. We call such braids equivalent. It
was shown by Markov \REF\Markov{A.A. Markov, Recueil Math. Moscou 1 (1935) 73}
\r\Markov, that two braids are equivalent if an only if they
can be related by the sequence of moves of the two types,
$$({\rm I})\qquad AB\rarrow BA \qquad {\rm for\ } A,B\in B_n,\e$$
$$({\rm II})\qquad A\rarrow A G_n^{\pm 1} \qquad {\rm for\ }
A\in B_n,\e$$
where $B_n$ denotes the braid group on $n$ elements, defined by the relations
eq. (17).

In order to classify links we wish to form a functional $\alpha$ which
assigns a complex number for each link, in such a way that topologically
equivalent links will have the same value of $\alpha$,
$\alpha(A)=\alpha(B)$ if $A$ and $B$ are equivalent topologically.
To do so, it is thus sufficient to demand that $\alpha$ is invariant
under the Markov moves. We define a Markov trace on a braid, $\phi(A)$,
for $A\in B_n$, to be a complex number obeying the
properties,
$$\eqalign{({\rm (I)}\qquad \phi(AB)&=\phi(BA),\qquad A,B\in B_n,\cr
({\rm II})\qquad \phi(A G_n)&=\tau \phi(A), \qquad \phi(A G_n^{-1})=
\bar\tau \phi(A), \qquad A\in B_n,\cr}\e$$
and where the parameters $\tau$ and $\bar \tau$ are
$$\tau=\phi(G_i),\qquad \bar \tau=\phi(G_i^{-1}).\e$$
The link invariant $\alpha(A)$ is formed in terms of the Markov trace
$\phi(A)$, by
$$\alpha(A)=(\tau\bar\tau)^{-(n-1)/2} (\tau/\bar \tau)^{e(A)/2} \phi(A),\e$$
where $e(A)$ is the exponent sum of the braid, i.e.,
$$e(\prod_{i=1}^n G_i^{a_i})=\sum_{i=1}^n a_i, \e$$
which is evidently a well defined grading, since it is preserved by the braid
group relationships, eqs. (17).

We next proceed to describe a Markov trace based on the lattice model
IRF$({\cal O},x)$. Note that any element of the braid group, $A\in B_n$
is represented by some diagonal to diagonal transfer matrix,
$A_{l_1,l_2,\ldots,l_n}^{m_1,m_2,\ldots, m_n}$, where the generators
are represented by the conformal braiding matrix $G_i$. Now, define the
diagonal matrix,
$$(H^n)_{l_1,l_2,\ldots l_n}^{m_1,m_2,\ldots, m_n}=
\prod_{i=1}^n \delta(l_i,m_i) {S_{l_n,0}\over S_{l_1,0}},\e$$
where $S$ is, as before, the torus modular matrix, which gives the
crossing multiplier. Define also a constrained trace by,
\def\htr{\mathop{\hat{\rm Tr}}\nolimits}
$$\htr(A)=\sum_{l_2,l_3,\ldots,l_n} A_{l_1,l_2,\ldots,l_n}^{l_1,l_2,
\ldots,l_n}.\e$$
Then the Markov trace is defined by
$$\phi(A)={\htr(H^n A) \over \htr(H^n)},\e$$
for any element of the braid group $A$.
It remains to show that the Markov trace so defined, $\phi(A)$ obeys
the properties (I) and (II), eqs. (18--19). Property (I) follows trivially
from the definition, while property (II) follows from a straight forwards
calculation, provided that the Boltzmann weights obey the Markov property,
$$\sum_{b\equiv a} \bw b a a c u {S_{b,0}\over S_{a,0}}=H(u)\rho(u),\e$$
where $H(u)$ is some function independent of $a$ and $c$.
The parameters $\tau$ and $\bar \tau$ are given by
$$\tau,\bar\tau=\lim_{u\rarrow\pm\infty} H(u)/H(0),\e$$
where $\tau$ ($\bar\tau$) corresponds to the plus (minus) sign in the limit.

Using the crossing property, eq. (12), it is straight forwards to show that the
extended Markov property holds provided that the following relation
is valid,
$$X_i(\lambda) X_i(u)=\beta(u) X_i(\lambda),\e$$
where $\beta(\lambda-u)=H(u)\rho(u)$.
We shall now show that for the models IRF(${\cal O},x)$ the property eq. (29)
holds and that thus $\phi$ is always a good Markov trace.
This is a simple calculation using eqs. (8). We note that
$X_i(\lambda)=P^N_i f^N(\lambda)$, since $f^a(\lambda)$ vanishes for $a\neq N$.
Thus $X_i(\lambda)$ is indeed a projection operator and so
$$X_i(\lambda) X_i(u)=f^N(\lambda) \sum_{a=1}^N P^N_i P^a_i f^a(u)=
\beta(u) X_i(\lambda),\e$$
where we used eqs. (3,8), and
$$\beta(u)=f^N(u)=\prod_{a=1}^N \sin(\zeta_i+u).\e$$
It follows that the parameters are
$$H(u)=\prod_{i=1}^N {\sin(\lambda+\zeta_i-u)\over \sin(\zeta_i+u) },\e$$
and
$$\tau=e^{iN\lambda}\prod_{i=1}^N {\sin(\zeta_i)\over \sin(\lambda+\zeta_i)},
\e$$
and $\bar\tau=\tau^\dagger$.
It follows that the invariant we defined, eq. (22), indeed assumes the
same values for topologically equivalent links, and thus can be used
to classify knots and links.

For a number of examples of IRF models, the link invariants we defined
here were
previously calculated (for a review, see \r\Wadati, and references therein).
For example, the unrestricted Lie algebra model $A_{m-1}$ give rise
to the HOMFLY polynomial \REF\HOMFLY{P. Freyd, D. Yetter, J. Hoste, W.B.R.
Lickorish, K. Millet and A. Ocneanu, Bull. Am. Math. Soc. 12 (1985) 239}
\r\HOMFLY\ (as a polynomial in $m$ and the crossing parameter), which
is a two variable generalization of the original Alexander
polynomial \REF\Alex{J.W. Alexander, Trans. Am. Math. Soc. 30 (1928) 275}
\r\Alex\ (at the limit $m\rarrow 0$) and the more recent Jones polynomial
\REF\Jones{V.F.R. Jones, Bull. Am. Math. Soc. 12 (1985) 103}\r\Jones\
($m=2$ case). The unrestricted $B_m$, $C_m$ and $D_m$ IRF models
give the Kauffman polynomial \REF\Kauf{L.H. Kauffman, Trans. Am. Math. Soc.,
to be published.}\r\Kauf.
These models correspond to the current algebra RCFT based on the Lie algebras
$A$, $B$, $C$, $D$, with the field which is the fundamental representation
for $A_n$, and the vector representation for the other algebras.

It is noteworthy that the construction presented
here, while encompassing all the known link invariants, provides for a
very far reaching generalization of these, along with a unified framework
for their construction. Such new invariants are indeed needed in the
problem of classifying links as it is well known that two topological distinct
links may certainly have identical classifying polynomials (see for example
Birman's example \REF\Birman{J.S. Birman, Invent. Math. 81 (1985) 287}
\r\Birman\ of two different knots that have the same Jones polynomial).

The link invariants we defined eq. (22) may be calculated directly for each
IRF model by substituting the Boltzmann weights and preforming
the traces. This is however rather cumbersome for big links. A considerable
simplification
is provided by the skein relations which relate the invariants of different
links
\REF\Conway{J.H. Conway, in Computational Problems in Abstract Algebra, ed. J.
Leech (Pargamon, New--York, 1970) p. 329}\r{\Alex,\Conway}.
To derive skein relations for the invariants described here, first
note that the Braiding matrix $G_i$ obeys a fixed $N$th order polynomial
equation,
$$\sum_{m=0}^N a_m G_i^m=\prod_{m=1}^N (G_i^m-\lambda_m)=0,\e$$
where we used eq. (5).
Define the link $L_{m}$ to be the link obtained with the insertion
of the braid element $G_i^m$, i.e., if $L$ described by the braid
$A$, then $L_m$ is described by the braid $AG_i^m$. Using the polynomial
relation, eq. (34), we find immediately the relation for the Markov trace,
$$\sum_{m=-k}^{N-k} a_m \phi(L_m)=0,\e$$
for any $k$. Substituting this into the definition of the invariant, eq. (22),
we find the skein relation,
$$\sum_{-k}^{N-k} b_m \alpha(L_m)=0,\e$$
where
$$b_m=a_m (\tau/\bar \tau)^{-m/2}.\e$$
The skein relation, eq. (36), is a very effective tool for the calculation of
link invariants.

We thus describe in this note a whole wealth of link invariants which are
in a one to one correspondence with a pair of a rational conformal field
theory and a field in this theory. The RCFT and the field chosen
are arbitrary, and every RCFT gives rise to different invariants.
It is tantalizing to speculate on this in a number of directions.
First, since all known link polynomials arise in this fashion, one might
conjecture that the category of link invariants and the category
of pairs of conformal field theory and a field in it are in fact equivalent
ones, and that the problem of classifying link invariants is thus the same
as that of classifying conformal field theory. Second, one might
ponder the generalization of these ideas to all conformal
field theories, not necessarily rational. There does not seem to be any
obstacle in doing so, and the entire construction might be carried,
mutatis mutandis.
This will also open up an entire
different type of invariants, so called irrational, which, in particular,
obey an infinite order skein relations, i.e., a Laurent series type rather
than polynomial. Such invariants appear not to have been studied before.

Finally, it is hoped that the results described here will be of help
in the further understanding of both knot theory and two dimensional
physics, along with the fascinating interrelationship between them.
\refout
\bye